\documentstyle[preprint,aps]{revtex}
\newcommand{\beq}{\begin{eqnarray}}
\newcommand{\eeq}{\end{eqnarray}}
\begin{document}
\draft

\title
{Size Dependence In The Disordered Kondo Problem}
\author{Ivar Martin, Yi Wan, and Philip Phillips}
\vspace{.05in}

%
\address
{Loomis Laboratory of Physics\\
University of Illinois at Urbana-Champaign\\
1100 W.Green St., Urbana, IL, 61801-3080}

%
\maketitle

\begin{abstract}
We study here the role randomly-placed non-magnetic scatterers play on the
Kondo effect.  We show that spin relaxation 
effects (with time $\tau_s^o$) in the vertex corrections
to the Kondo self-energy lead to an exact cancellation
of the singular temperature dependence arising from the
diffusion poles.
For a thin film of thickness $L$ and a mean-free path
$\ell$, disorder provides a correction to the 
Kondo resistivity of the form $\tau_s^o/(k_FL\ell^2)\ln T$ that explains both the 
disorder and sample-size depression of the Kondo effect observed by
Blachly and Giordano (PRB {\bf 51}, 12537 (1995)).
\end{abstract}

\pacs{PACS numbers:72.10.Fk, 72.15.Nj, 75.20.Hr} 

\narrowtext
At low temperatures, the resistivity of a metal alloy acquires a logarithmic 
temperature dependence\cite{kondo} in response to spin-flip scattering between local magnetic impurities and the conduction
electrons.  This behaviour persists down to a temperature (the Kondo temperature, $T_k$)
where the magnetic impurities and conduction electrons
begin to condense into singlet states.  While perturbation theory is sufficient to 
establish the existence of 
the $\ln T$ term, its presence ultimately signifies that perturbation theory
is breaking down.  Alternatively, spin-flip scattering between 
conduction electrons and localized 
magnetic centers has a singular frequency ($\omega$)
dependence. Magnetic impurities are not alone in this respect.  It is well-known
that even non-magnetic impurities can generate a singular ($\ln\omega$ in d=2)
frequency dependence in the conductivity\cite{go4}.  In a sample containing
both magnetic and non-magnetic impurities, the question arises: which singularity
ultimately wins or can the interplay between the singularities lead to a suppression
of either localization or the Kondo effect?  In this letter, we resolve these questions.

The motivation for this study is two-fold.  First, while there have been numerous
treatments of this problem\cite{everts}-\cite{vz}, a clear consensus has not been reached.  To illustrate,
Everts and Keller\cite{everts} were first to
show that non-magnetic scattering contributes a $1/\sqrt{T}$ to the Kondo
self-energy that dominates
the Kondo $\ln T$ singularity at low temperatures in d=3.  
Bohnen and Fisher\cite{bf} argued, however, that such a term would not survive
in the conductivity.  More recently,
Ohkawa, Fukuyama, and Yosida\cite{ohk} showed
that disorder results in a singularity of the form $T^{d/2-2}$ in the conductivity.
At low temperatures this singularity dominates the Kondo $\ln T$.  As a result,
these groups conclude that static disorder can mask the Kondo resistivity as
$T\rightarrow 0$.  On the experimental side, Blachly and Giordano\cite{bg}
recently measured the conductivity in a series of thin films containing magnetic
as well as non-magnetic impurities.  They found no evidence for the 
$T^{d/2-2}$ singularity but observed instead a suppression of the Kondo resistivity
as the strength of the disorder increased.  Earlier experiments by Korn\cite{korn} also failed
to observe the $T^{d/2-2}$ singularity but observed instead an enhancement in the 
Kondo resistivity.  The point of agreement between these experiments is that disorder
couples non-trivially to the Kondo effect and ultimately modifies 
the coefficient of the $\ln T$ dependence. Given the strong dimensional dependence of the localization transition,
disorder could eventually lead to a sample size dependence of the Kondo effect.

At the outset, we set aside the still controversial 
issue (ref. 8c) of the sample size dependence
and focus on the seemingly straightforward problem of the role non-magnetic
disorder plays in the Kondo effect.  The new wrinkle we introduce in this problem
is the feedback effect spin scattering has on localization.  While it is 
standard to consider the direct influence of localization on the Kondo effect,
the reverse effect has not been included\cite{footnote}.  Nonetheless, it is well-known that
electron scattering by disordered Heisenberg spins introduces a cutoff of the
diffusion pole in both the particle-hole (diffuson) and particle-particle (Cooperon)
channels except for the $S=0$ particle-hole channel\cite{altshuler}.  
When spin-scattering is included in the diffusion propagators, 
the fate of the $T^{d/2-2}$ singularity rests
on whether the $S=0$ particle-hole propagator contributes to the Kondo self-energy.
We show explicitly it {\it does not}.  

The starting point for our analysis is a model Hamiltonian $H=H_o+H_{sd}$ 
that contains both normal impurities
\begin{equation}
H_o=\sum_{k\sigma}\left(\varepsilon_k-\varepsilon_F\right)a_{k\sigma}^{\dagger}a_{k\sigma}
+{\frac v\Omega}\sum_{k,k',i}e^{\bf i(k-k^{\prime})\cdot R_i}
a^{\dagger}_{k\sigma} a_{k'\sigma}
\end{equation}
as well as magnetic scatterers
\beq
H_{sd}=-{\frac J \Omega}\sum_{R_n,k,k',\sigma,\sigma'}
e^{\bf i(k-k^{\prime})\cdot R_n}{\bf \sigma}_{\sigma,\sigma'}\cdot {\bf S_n}
a_{k\sigma}^{\dagger}a_{k'\sigma'}.
\eeq
The operator $a_{k\sigma}^{\dagger}$ creates an electron in a plane wave state with momentum
$k$ and spin $\sigma$ and energy $\varepsilon_k$, $v$ measures the strength of the 
scattering with the non-magnetic disorder, $J$ is the exchange interaction, $R_n$ 
denotes the position of the impurities, magnetic or otherwise, ${\bf S}_n$ is the spin
operator for the magnetic impurity at site $n$, ${\bf \sigma}$ is the Pauli spin
operator, and $\Omega$ is the volume.  The two natural timescales in this problem
are, $\tau_s^o$ and $\tau_o$, the magnetic and non-magnetic scattering times.
In terms of the density of states of the host metal, $\rho_o$ and the 
concentrations of magnetic and non-magnetic scatterers, $n_s$ and $n_o$, respectively,
we have that $\hbar/2\tau^o_s=3\pi n_s\rho_o |J|^2/4$ and 
$\hbar/2\tau_o=\pi n_o\rho_o |v|^2$. The total scattering rate
is $1/\tau=1/\tau^o_s+1/\tau_o$.   To measure the strength of the non-magnetic
disorder, we define $\lambda=\hbar/(2\pi\varepsilon_F\tau_o)$.  We assume that the concentration
of localized spins is dilute so that long-range spin glass effects are irrelevant.
Also, we work in the regime in which normal impurity scattering dominates, $1/\tau_o\gg 1/\tau_s^o$

To evaluate the conductivity above $T_k$, we must first calculate the Kondo
self energy.  To include the dynamical effects of the localized spins, it is
sufficient to calculate the self energy to third order in the exchange interaction
$J$.  At this order, static disorder can be  included by decorating the single
and double spin-flip
vertices with Cooperon and diffuson propagators\cite{ohk},\cite{vz}.  
In previous work\cite{ohk},{\cite{vz}, spin-independent Cooperons and 
diffusons of the form
$C(Q,\omega)=D(Q,\omega)\propto \frac {1}{(DQ^2-i\omega)}$ were used
where $Q$ and $\omega$ are the net momentum and energy transfer and
$D=2\hbar\varepsilon_F\tau/dm$ is the diffusion constant.
However, this is inconsistent because $C(Q,\omega)$ and $D(Q,\omega)$ are
coupled to electron lines of different spin.  Such 
propagators are well-known\cite{altshuler} to depend on spin
and hence we include explicitly the spin dependence here.
If all scattering processes are 
treated in the first Born approximation, 
the Cooperon propagator\cite{altshuler} is transformed
to
\beq
C_{\alpha\beta\gamma\delta}&=&\frac {\hbar^2}{8\pi\rho_o\tau^2 (DQ^2-i\omega+2/\tau_s^o)}
\left(\delta_{\alpha\beta}\delta_{\gamma\delta}-\sigma_{\alpha\beta}\cdot\sigma_{\gamma\delta}\right)
\nonumber\\
&&+\frac {\hbar^2}{8\pi\rho_o\tau^2(DQ^2-i\omega+2/3\tau_s^o)}
\left(3\delta_{\alpha\beta}\delta_{\gamma\delta}+\sigma_{\alpha\beta}\cdot\sigma_{\gamma\delta}\right)
\eeq
and the diffuson becomes
\beq\label{eq:dif}
D_{\alpha\beta\gamma\delta}&=&\frac {\hbar^2}{8\pi\rho_o\tau^2 (DQ^2-i\omega)}
\left(\delta_{\alpha\beta}\delta_{\gamma\delta}+\sigma_{\alpha\beta}\cdot\sigma_{\gamma\delta}\right)
\nonumber\\
&&+\frac{\hbar^2}{8\pi\rho_o\tau^2 (DQ^2-i\omega+4/3\tau_s^o)}\left(3\delta_{\alpha\beta}
\delta_{\gamma\delta}-\sigma_{\alpha\beta}\cdot\sigma_{\gamma\delta}\right).
\eeq
The survival of the diffusion pole in the spin-
dependent diffuson is a consequence of particle-hole conservation. The two terms
in each of these propagators correspond to singlet and triplet scattering, respectively.
Pairs
of spin indices $\alpha\beta$ and $\gamma\delta$ are indexed chronologically
along the particle lines that comprise the diffusion ladder diagrams. 

The diagrams shown in Fig. 1a contain the dominant quantum corrections to the Kondo
self-energy at third order in the presence of disorder. The sum of all such diagrams
is 
\beq\label{eq:self}
\Sigma_{3q}(k,i\epsilon_n)&=&\frac{2}{\beta^2}\sum_{\omega_\ell,\omega_m,Q,q,\pm}
V_{\alpha\beta\nu\eta}^\pm(i\omega_\ell,i\omega_m)
G(i\epsilon_n+i\omega_m,q)\times\nonumber\\
&&\left[G(i\epsilon_n+i\omega_\ell,k+Q)+n_o|v|^2\sum_{k^\prime}G^2(i\epsilon_n,k^\prime)
G(i\epsilon_n+i\omega_\ell,k^\prime+Q)\right]\times\nonumber\\
&&(D_{\sigma\alpha\beta\gamma}(i\omega_\ell,Q)
D_{\gamma\nu\eta\sigma}(i\omega_\ell,Q)+C_{\sigma\alpha\gamma\nu}(i\omega_\ell,Q)
C_{\beta\gamma\eta\sigma}(i\omega_\ell,Q))
\eeq
where $G(i\epsilon,q)$ is the electron Green function
\beq
G(i\epsilon,q)=\frac{1}{i\epsilon+\epsilon_F-\hbar^2q^2/2m+i(\hbar/2\tau)sgn(\epsilon)},
\eeq
the electron energies are
the Matsubara frequencies, $\epsilon_n=(2n+1)\pi T$, the psuedofermion
energies are $z_k=(2k+1)\pi T$, $\omega_\ell=2l\pi T$, 
$DQ^2<\hbar/\tau_o$ and $(\epsilon_n+\omega_\ell)\omega_\ell<0$. We
have set $k_B=1$.
The factor of 2 arises from the two possible couplings of the diffusion
progagators to the internal electron lines and the $\pm$ from the two orientations of the
psuedofermion loops.
The psuedofermion part involves a trace over the components of the
impurity spin operators
and hence simplifies to
\beq
V_{\alpha\beta\nu\eta}^\pm(i\omega_l,i\omega_m)=\frac{1}{4}J^3n_s\beta
\left[\frac{1}{i\omega_\ell}(\delta_{m0}-\delta_{\ell m})
(1-\delta_{\ell 0})
+\frac{1}{i\omega_m}\delta_{\ell 0}(1-\delta_{m0})\pm\frac{\beta}{2}\delta_{m0}\delta_{\ell 0}
\right]
(\sigma_{\alpha\beta}^a\sigma_{\nu\eta}^a)
\eeq

From the psuedofermion contribution, we see that the sum over
the spin indices separates into two identical sums of the form,
$\sum_{\alpha\beta}D_{\sigma\alpha\beta\gamma}
\sigma_{\alpha\beta}^a$.
If we use the identity
$\sum_{\alpha\beta}(\sigma_{\nu\alpha}\cdot\sigma_{\beta\gamma})\cdot
\sigma_{\alpha\beta}^a = -\sigma_{\nu\gamma}^a$,
we find immediately that the cancellation of the divergent diffusion terms
\beq\label{cancel}  
\sum_{\alpha\beta}D^{S=0}_{\nu\alpha\beta\gamma}\sigma_{\alpha\beta}^a
&\propto& \sum_{\alpha\beta}(\delta_{\nu\alpha}
\delta_{\beta\gamma}+\sigma_{\nu\alpha}\cdot\sigma_{\beta\gamma})
\sigma_{\alpha\beta}^a=0
\eeq
from the $3^{rd}$ order Kondo self-energy is exact. To any order in $J$, the cancellation
of the diffusion pole can be seen as follows.  In the most 
divergent approximation, each diffuson encircles a vertex that is exactly
equal to the Abrikosov\cite{abrikosov} vertex function 
$\Gamma\propto\sigma\cdot {\bf S}$.  When this function
is now multiplied by $D^{S=0}$ and summed over the spin indices,
the cancellation to all orders follows immediately from Eq. (\ref{cancel}).
This is one of the principal results of this
paper.  The cancellation of the $S=0$ component of the diffuson 
is fundamentally tied to the fact that the Kondo
interaction does not conserve spin.  Further, it signifies
that the resultant conductivity is independent of the pure charge density propagator.
Summing over the spin indices in the remaining propagators in the self energy 
reduces the problem to one in which the diffuson and Cooperon 
are spin independent:
$\tilde{D}=\hbar^2/(2\pi\rho_o\tau^2)(DQ^2-i\omega+4/3\tau_s^o)^{-1}$ and
$\tilde{C}=\hbar^2/(4\pi\rho_o\tau^2)[(DQ^2-i\omega+2/\tau_s^o)^{-1}+
(DQ^2-i\omega+2/3\tau_s^o)^{-1}]$.  When $1/\tau_s^o=0$ (or
equivalently, $T\gg \hbar/\tau_s^o$), we recover
the standard form for these propagators. 

To calculate the resistivity, we evaluate the standard self-energy
as well as the Cooperon weak-localization diagrams\cite{ohk}. Because the results of the calculation are 
rather lengthy, we present here only the asymptotic behaviour.  
In the limit
$T\gg \hbar/\tau^o_s$ (as in the case when the magnetic impurities are
dilute), we recover the inverse temperature dependence
\beq\label{hight}
\frac{\hbar}{2\tau^{C}}=\frac{\hbar}{2\tau^{D}}=
\frac{-\pi\hbar\rho_o\lambda J}{3\tau_o}\frac{\hbar}{\tau_s^o T}\ll-\rho_o\lambda J\frac{\hbar}{\tau_o}
\eeq
of refs. (\cite{vz}, \cite{ohk}).  In the opposite regime,
$T\ll\hbar/\tau^o_s$, the limiting forms of the Cooperon and diffuson relaxation times
\beq\label{lowt1}
\frac{\hbar}{2\tau^{D}}+\frac{\hbar}{2\tau^{C}}&=&-\left(\frac52+\frac{3\ln 3}{4}\right)\rho_o\lambda J
\frac{\hbar}{\tau_o}\ln\frac{\hbar}{T\tau_s^o}
\eeq
are both logarithmic functions of temperature.   

The final contribution to the relaxation time comes from the 
the Cooperon weak-localization
diagram.  In two dimensions in the presence of spin-flip
scattering, the weak-localization contribution is
$\Delta\sigma_{loc}=-e^2/(2\pi^2\hbar)\ln (\sqrt 3\tau_{\epsilon s}/\tau_o)$,
where $\hbar/2\tau_{\epsilon s}=8\hbar/(3\tau_s^o)(1-\rho_oJ\ln(\epsilon_F/T))$.
Physically, $\tau_{\epsilon s}$ plays the role of the inelastic scattering
time in the weak-localization correction.  Inclusion of the $3^{rd}$ order 
correction
to the spin scattering time, enhances the spin-flip scattering rate, thereby
weakening the effects of localization.   
To see this more clearly, we expand the argument of the logarithm
for temperatures well above the Kondo temperature:  
\beq\label{loc}
\Delta\sigma_{loc}=\frac{-e^2}{2\pi^2\hbar}\ln\left(\frac{3\sqrt 3\tau_s^o}{8\tau_o}\right)
-\frac{e^2}{2\pi^2\hbar}\rho_oJ\ln \left(\frac{\epsilon_F}{T}\right)
\eeq
We see clearly that the Kondo interaction reduces the weak localization correction because
$J<0$.   

We collect all the contributions discussed above to determine the conductivity.
In the temperature range $T_k\ll T<\hbar/\tau_s^o$, Cooperon, diffuson, and weak-localization
corrections are logarithmic in temperature. Combining the results from Eq. 
(\ref{lowt1}) with the weak-localization correction, we find 
that the magnitude of the logarithmic part of the conductivity 
\beq
\Delta\sigma^T=\sigma_o\frac{4\tau_o\rho_o J}{\tau_s^o}\left(1+1.4\lambda\frac{\tau_s^o}
{\tau_o}\right)\ln\frac{\epsilon_F}{T}
\eeq
is enhanced by disorder. The first term in this expression arises from the unperturbed
Kondo effect and the latter from the interplay with disorder.  
Inclusion of disorder in the self energy always enhances the Kondo resistivity
by increasing repetitive scattering
at magnetic impurities. 

For temperatures $T\gg \hbar/\tau_s^o$, the self-energy
contribution to the relaxation time scales as $1/T$, whereas the 
weak-localization
correction is proportional to $\ln T$. However, comparison of the magnitude
of these corrections (see Eqs. (\ref{hight}) and (\ref{loc})) 
reveals that the weak-localization term dominates and the magnitude of
the resultant logarithmic correction
\beq\label{supps}
\Delta\sigma^T=\sigma_o\frac{4\tau_o\rho_o J}{\tau_s^o}\left(1-
\frac{\lambda\tau_s^o}
{4\tau_o}\right)\ln \frac{\epsilon_F}{T}
\eeq
is suppressed by the disorder. The ratio $\lambda/\tau_o$ scales as
$1/\ell^2$, where $\ell$ is the mean-free path.  We see then that in the dilute
impurity regime, disorder suppresses the Kondo effect.  The crossover from
enhancement to suppression of the Kondo effect occurs
because the magnitude and functional dependence of the quantum corrections
to the self-energy are
determined by the shortest of two length scales: the phase-breaking
length, $L_\phi=\sqrt{D\tau_s^o/\hbar}$
and the diffusion length,
$L_T=\sqrt{D/T}$.  The latter arises because coupling of diffusion propagators
to internal electron lines in the self-energy leads to an effective electron-electron interaction.

Let us now apply our results to the experiments on thin films by
Blachly and Giordano\cite{bg}.  In all of their samples 
the film thickness $L$ satisfied the inequality $\ell<L\ll L_\phi$.
Hence, we can treat the films as quasi-2d with respect to localization,
but because $\ell<L$ the electron gas is characterized by a 3-dimensional density
of states $\rho_o=1/(2\pi)^2(2m/\hbar^2)^{3/2}\epsilon^{1/2}$ with a diffusion
constant given by $D=2\hbar\epsilon_F\tau_o/3m$. The summation on $Q$
in the Cooperon and diffuson is restricted to small momentum transfers such
that $DQ^2<1/\tau_o$. However, for thicknesses of the sample
on the order of $\ell$, the smallest wave vector in the transverse direction
does not satisfy this constraint.  To rectify this problem, Volkov\cite{volkov}
showed
that surface
boundary conditions must be treated consistently. For thin films, his treatment
shows that the boundaries always give rise to a strictly 2-dimensional 
weak-localization
correction and an explicit
finite size dependence. To account for the former, the momentum integration 
in the Cooperon and diffuson must be restricted to the plane.
The density of states
that arises from converting the sum to an integral will be the 2-dimensional
density of states $\rho_o^{2D}=\pi\rho_o/(k_FL)$.  Hence, the self-energy
diagrams will generate a size-dependence to the conductivity.
The explicit finite-size weak-localization correction is\cite{volkov} 
$\Delta\sigma_{loc}=-e^2/(2\pi^2\hbar L)\ln 
\left(\sqrt 3\tau_{\epsilon s}/\tau_o\left(\sinh(L/\ell)(\ell/L)\right)\right)$.
The size-dependence in the logarithm yields an effective size dependence in
the spin-relaxation time.  However, this will not affect the temperature dependence
of the conductivity.  Hence, the only size dependence that is coupled to the
temperature is the $1/L$ prefactor of the weak-localization correction. 

We now combine these results in the low and high-temperature limits discussed
earlier.  In the two limits, we obtain
\begin{equation}\label{final}
\Delta\sigma^T=\left\{\begin{array}{ll}
\sigma_o\frac{4\tau_o\rho_o J}{\tau_s^o}\left(1+
\frac{2.3\hbar\tau_s^o}
{\pi mk_FL\ell^2}\right)\ln \frac{\epsilon_F}{T}&{\rm if}\quad
T_k\ll T<\hbar/\tau_s^o\\ 
\sigma_o\frac{4\tau_o\rho_o J}{\tau_s^o}\left(1-
\frac{1.2\hbar\tau_s^o}
{\pi mk_FL\ell^2}\right)\ln \frac{\epsilon_F}{T}&{\rm if}\quad
T_k,\hbar/\tau_s^o\ll T \end{array}\right.
\end{equation}
an explicit size and disorder correction that scales as $1/(\ell^2L)$.  In
the concentrated impurity limit $T<\hbar/\tau_s^o$, increasing disorder
enhances the resistivity. In Cu(Fe) alloys at impurity 
concentrations ranging from $0.3-2.1\%$, Korn observed an enhancement in the Kondo 
resistivity that is consistent with the first equation above.  However, in the dilute
limit, $T\gg\hbar/\tau_s^o$, we predict a suppression of the Kondo 
effect as the disorder is increased and the size of the sample decreases.
In the experiments of Blachly and Giordano\cite{bg}, $\hbar/\tau_s^o\approx 0.1K$
which is much less than the Kondo temperature for $Cu(Fe)$. The second
of equations should be valid.  Fig. 2 shows a comparison between the experimental
data and the theoretical predictions. The best fit to the data
was obtained with $\tau_s^o=0.52ns$ which is consistent with the experimental
range of $10^{-10}s$.  As is evident, theory and experiment are in 
good agreement.
We also obtained quantitative agreement with the experimental data 
when the sample
size was varied. We conclude that disorder can suppress the Kondo
resistivity and give rise to a sample size dependence of the form $1/(\ell^2 L)$.
We note in closing that a recent theory of the Kondo size dependence in clean
samples has been proposed by Ujsaghy, Zawadowski, and Gyorffy\cite{zawa}.
This approach applies strictly in the ballistic case where $\ell\gg L$.

\acknowledgments
We thank Dan Ralph for making us aware of the Giordano experiments and 
Nick Giordano,
Yuri Lyanda-Geller and Eduardo Fradkin for useful
discussions on spin-scattering. This work is supported in part by the NSF 
grants No. DMR94-96134.

\begin{figure}
\caption{ Feynman diagrams contributing to the Kondo self-energy ($\Sigma$).
The dashed lines correspond to Abrikosov psuedofermions and the double solid
lines to diffusons and double dashed lines to the Cooperons.  The Greek
letters indicate the spin. The
$X$ indicates a single non-magnetic impurity scattering event.}
\label{fig1}
\end{figure}
\begin{figure}
\caption{Comparison of the theroretical
prediction for the Kondo resistivity predicted from the second
of Eq. (\protect\ref{final}) with the experimental data of Blachly and Giordano
(\protect\cite{bg}) Fig. 7.  The horizontal axis measures the strength of
the static disorder through the mean-free path.}
\label{fig2}
\end{figure}

\begin{thebibliography}{99}
\bibitem{footnote}
In ref. 8, spin relaxation is included in the Cooperon contribution to the 
conductivity.  However, the equivalent contribution in the self-energy was ignored.
\end{thebibliography}
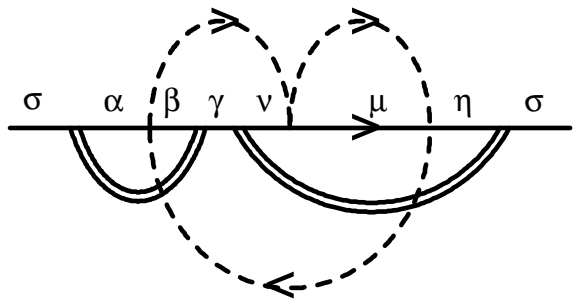
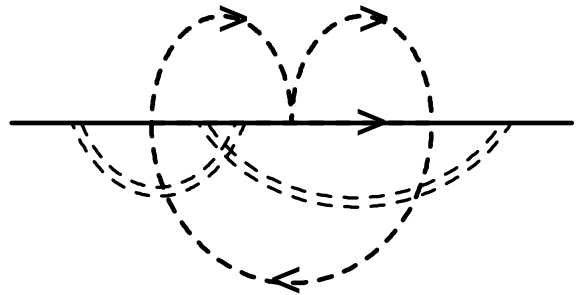
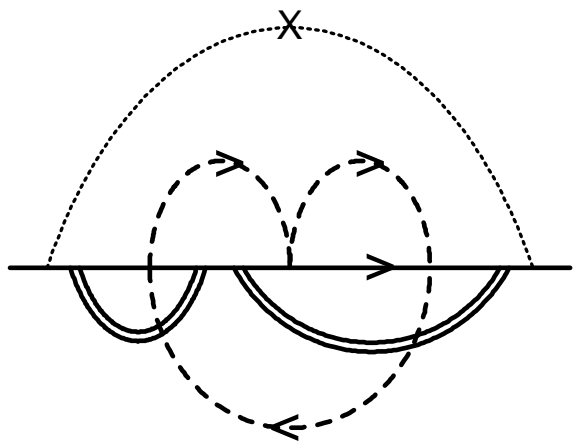
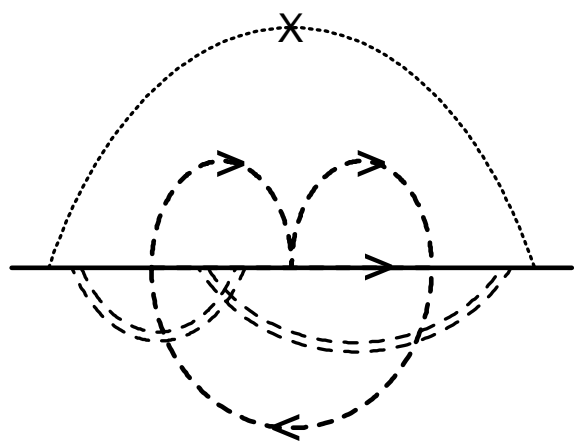
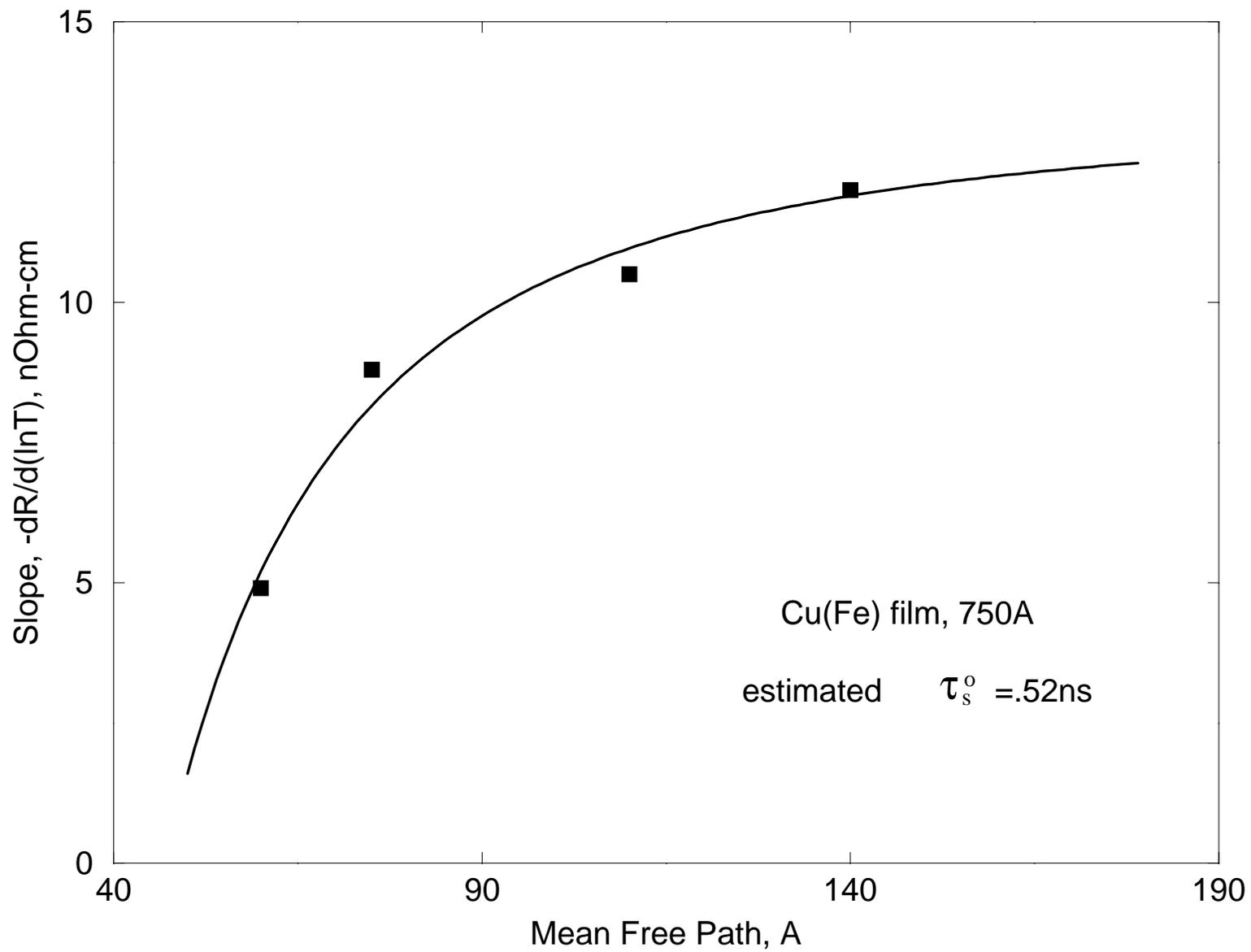
\end{document}